\documentclass[prl,reprint,showpacs,twocolumn,superscriptaddress,floatfix]{revtex4-1}

\usepackage{epsfig}
\usepackage{amsmath}
\usepackage{amssymb}
\usepackage{amsfonts}
\usepackage{mathptmx}
\usepackage{dcolumn}
\usepackage{eucal}
\usepackage{bm}
\usepackage{color}
\usepackage[colorlinks,linkcolor=blue,citecolor=blue]{hyperref}
\usepackage{epstopdf}
\usepackage{bbold}
\usepackage{url}
\usepackage{mathtools}

\usepackage{dsfont}

\newcommand{\Time}{\mathcal{T}}

\newcommand{\average}[1]{\langle #1 \rangle}
\newcommand{\Trace}{{\rm Tr}}

\newcommand{\re}{{\rm Re}}
\newcommand{\im}{{\rm Im}}

\def \ket#1{\mathinner{|{#1}\rangle}}

\newcommand{\matrixel}[3]{{\mathinner{\langle{#1}| {#2} | {#3}\rangle}} }

\newcommand{\GFuncF}{\mathcal{G}_{\chi,\mathcal{F}}}

\newcommand{\GFuncU}{\mathcal{G}_{\chi, \Delta U}}

\newcommand{\Prob}{\mathcal{P}}
\newcommand{\CNOT}[2]{{\rm CX}_{#1, #2}}

\newcommand{\Idoperator}{\mathds{1}}

\begin{document}

\title{Measurement of work and heat in the classical and quantum regimes}

\author{P. Solinas}
\affiliation{Dipartimento di Fisica, Universit\`a di Genova, via Dodecaneso 33, I-16146, Genova, Italy}
\affiliation{INFN - Sezione di Genova, via Dodecaneso 33, I-16146, Genova, Italy}
\author{M. Amico}
\thanks{Mirko Amico is currently employed at Q-CTRL inc.}
\affiliation{The Graduate School and University Center, The City University of New York, New York, NY 10016, USA}
\author{N. Zangh\`i}
\affiliation{Dipartimento di Fisica, Universit\`a di Genova, via Dodecaneso 33, I-16146, Genova, Italy}
\affiliation{INFN - Sezione di Genova, via Dodecaneso 33, I-16146, Genova, Italy}

\date{\today}

\begin{abstract}
Despite the increasing interest, the research field which studies the concepts of work and heat at quantum level has suffered from two main drawbacks: first, the difficulty to properly define and measure the work, heat and internal energy variation in a quantum system and, second, the lack of experiments.
Here, we report a full characterization of the dissipated heat, work and internal energy variation in a two-level quantum system interacting with an engineered environment.
We use the IBMQ quantum computer to implement the driven system's dynamics in a dissipative environment.
The 
experimental data allow us to construct quasi-probability distribution functions from which we recover the correct averages of work, heat and internal energy variation in the dissipative processes.
Interestingly, by increasing the environment coupling strength, we observe a reduction of the pure quantum features of the energy exchange processes that we interpret as the emergence of the classical limit.
This makes the present approach a privileged tool to study, understand and exploit quantum effects in energy exchanges. 
\end{abstract}

\maketitle

\section{Introduction}
\label{sec:Introduction}

The oddities of quantum mechanics such as particle entanglement, superposition of states, interference between evolution paths and so on, have proven to be a valuable asset to envision quantum devices able to outperform the corresponding classical ones.
From metrology \cite{Giovannetti2011}, to the detection of gravitational waves \cite{Aasi2013} passing through quantum computation and communication \cite{nielsen-chuang_book}, exploiting the quantum effects has given a decisive impulse towards new discoveries.
With the advent of quantum technologies this trend will be reinforced allowing for mass production of quantum-based devices. 
In this direction, studying the energy exchange processes of a quantum system with an external drive and an environment could have important implications for the future developments. This is the aim of a relatively new area of research referred to as quantum thermodynamics.

Although numerous results have been obtained, there is still no a clear consensus about how to determine what is the work and the heat in a quantum system \cite{ campisi2011colloquium, talkner2007fluctuation, dorner2013extracting, mazzola2013measuring, solinas2013work, solinas2015fulldistribution}.
In a closed quantum system, the work done on the system is equal to the {\it variation} of the internal energy of the system.
However, the need for information at different times makes it impossible to envision a proper measurement protocol for these quantities \cite{Perarnau-Llobet2017No-Go}.

This limitation has delayed experimental verification, especially in case of a quantum system undergoing dissipative dynamics.
Indeed, apart from a single experiment with closed quantum systems \cite{Batalho2014} and open ones \cite{pekola2013calorimetric, Gasparinetti2015, Viisanen_2015, Saira2016, Vesterinen2017, Ronzani2018, Peterson2019, Karimi2020, Karimi2020NatCom}, a full and convincing measurement of the the dissipated heat is still missing.
In this article, we fill this gap by showing that it is possible to obtain the correct and expected average values of work, heat and internal energy variation, and important information about the underlying quantum processes.

To avoid any confusion and interpretative pitfalls, we choose a more practical approach  by focusing on simple and precise questions.
Given a quantum system controlled by an external time-dependent field, how much energy does the system absorb during the evolution? How much work does the external field do on the quantum system? And what is the heat dissipated by the quantum system?

To answer to these questions, we implement the detection scheme proposed in Refs. \cite{solinas2013work,solinas2015fulldistribution,solinas2016probing} on a IBMQ device \cite{IBM_docs}.
In particular, we study a two-level quantum system, i.e., a qubit, subject to an external driving field and interacting with an engineered environment. The quantum detector and the engineered environment are represented by three additional qubits. The advantage of using an engineered environment is that we can tune the system-environment coupling strength and explore different dissipative regimes.
We implement the scheme on real physical qubits made available in the IBM quantum experience initiative.

Our physical observable is the phase of the detector qubit that is measured with standard techniques \cite{SM, IBM_docs} from which we recover the information about the average work, heat and internal energy variation while preserving the full quantum features of the evolution \cite{solinas2013work,solinas2015fulldistribution,solinas2016probing, SM}.

Furthermore, from the measured detector phase we are able to construct a quasi-characteristic generating function and a quasi-probability density function (QPDF) for these physical observables.
The QPDF reproduces the statistics of the two-measurement protocol (TMP) \cite{campisi2011colloquium} when the system is initially in an eigenstate of the Hamiltonian and keeps much more information about the evolution of the system.
In a direct analogy with the Wigner function \cite{WignerPhysRev1932}, the negative regions of the derived QPDF are associated to the violation of the Leggett-Garg inequalities and are the signature of a pure quantum phenomenon \cite{Leggett1985, solinas2016probing, clerk2011full, Lostaglio2018, Levy2020}.
The disappearance of these regions in presence of strong dissipation can be seen as a proof of the emergence of the classical limit in energy exchange processes induced by the presence of an environment.

\begin{figure}
    \begin{center}
    \includegraphics[scale=.4]{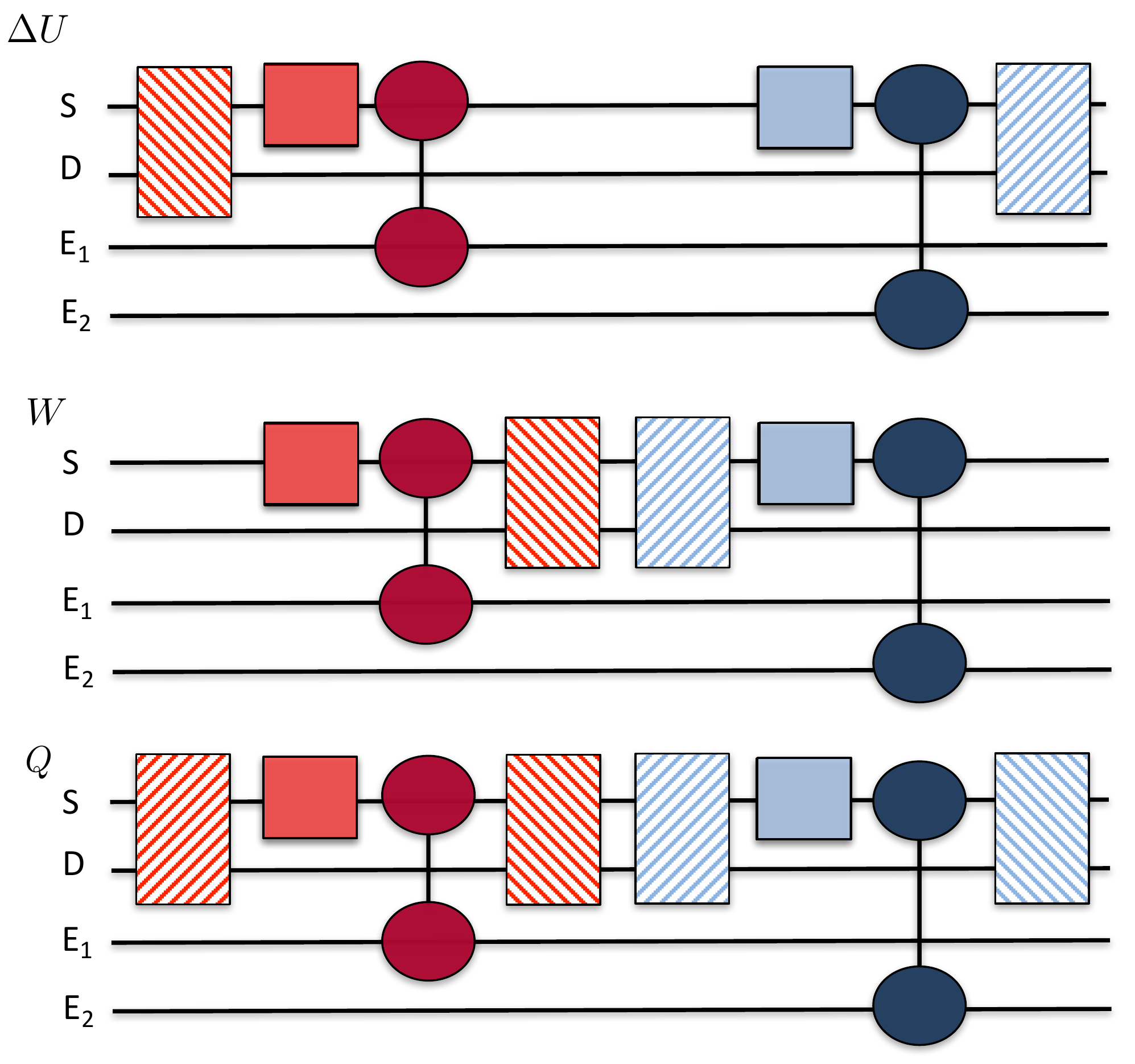}
       \end{center}
    \caption{Three schemes for the detection of the variation of internal energy $\Delta U$, work $W$ and dissipated heat $Q$. The letter S, D, E$_1$ and E$_2$ denote the system, the detector, the environment qubit $1$ and $2$, respectively.
The unitary dynamics are represented with red and light-blue filled squares corresponding to $U_{x}$ and $U_{z}$, respectively. The  dissipative gates are represented with circles connecting the S qubit with the E$_1$ and E$_2$ qubits. They corresponds to $R_x$ (dark red) and $R_z$ (dark blue). 
The system-detector coupling operations are represented with rectangles spanning the S-D qubit space.
They are $U_{- \chi, x}$ and $U_{ \chi, x}$ (see main text for the definition) represented with dashed red lines from up-left to down-right and up-right to down-left, respectively.
The operators $U_{- \chi, z}$ and $U_{ \chi, z}$ (see main text for the definition) are represented with dashed blue lines from up-left to down-right and up-right to down-left, respectively.
To simplify the presentation the gates needed to initialize the system and the detector are not shown (see \cite{SM} for details).
} 
    \label{fig:gate_sequence}
\end{figure} 

\section{System and dynamics description}
\label{sec:System_dynamics}

We start considering a two level system (denoted by $S$) that evolves under unitary evolution $U_S = U_z U_x$ with $U_z = \exp{(-i \beta \sigma_z)}$ and $U_x = \exp{(-i \alpha \sigma_x)}$ where $\sigma_i$ ($i=x, y, z$) are the usual Pauli operators.
The system is initially in the state $\ket{\psi_0} = \cos \frac{\theta}{2} \ket{0}_S+ \sin \frac{\theta}{2} e^{i \phi}\ket{1}_S$ where $\ket{0}_S$ and $\ket{1}_S$ are the eigenstates of $\sigma_z$.

This evolution is generated by the time-dependent Hamiltonian $H_S= \epsilon \sigma_x/2$ for $0 \leq t < t_1$ and $H_S= \epsilon  \sigma_z/2$ for $t_1 < t \leq \Time$ with appropriate $t_1$ and $\Time$.
The fact that the Hamiltonian changes in time assures that the external field does work on the system \cite{solinas2013work}.

The detector is represented by an additional two-level system (denoted by $D$). Its Hamiltonian is $H_D = \omega \Sigma_z/2$ (where the operators $\Sigma_i$ with $i=x, y, z$ are the Pauli operators acting on the detector) and it is time-independent.
The detector is initialized in an equal superposition of eigenstates of $H_D$, i.e., $(\ket{0}_D+ \ket{1}_D)/\sqrt{2}$.

The coupling Hamiltonian $H_{SD} = f(\chi, t) H_S(t) \otimes H_D$ allows us to store information about the system energy into the accumulated phase of the detector \cite{solinas2015fulldistribution,solinas2016probing}.
The function $f(\chi, t)$ in $H_{SD}$ determines the time at which the system-detector coupling is active and its coupling strength $\chi$.
If the system-detector coupling occurs on time scales much smaller than all of the other time-scales, we can assume that  $f(\chi, t_i) = \chi/\epsilon~\delta(t-t_i)$ which generates the transformation $U_{\chi, t_i} = \exp \{ i \frac{\chi}{\epsilon}~H_S(t_i) \otimes H_D\}$.
In particular, in relation the system dynamics described above, we have $U_{\pm \chi, x} = \exp \{\pm i \frac{\chi}{\epsilon}~\sigma_x \otimes H_D\}$ and $U_{\pm \chi, z} = \exp \{\pm i \frac{\chi}{\epsilon}~\sigma_z \otimes H_D\}$.

\begin{figure*}
    \begin{center}
    \includegraphics[scale=.6]{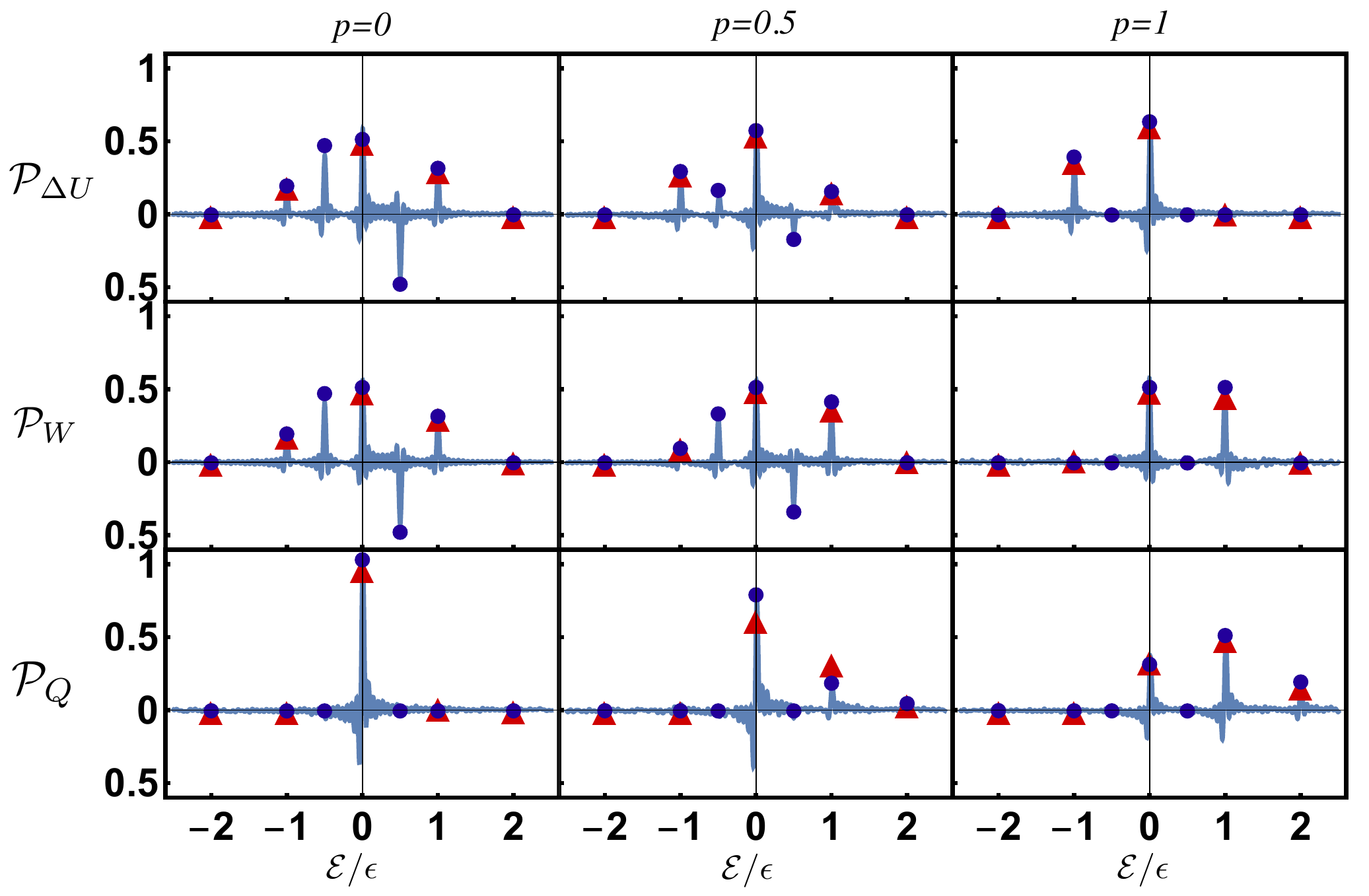}
       \end{center}
    \caption{
    The quasi-probability distribution functions for the variation of internal energy, work and dissipated heat (y-axis) with respect to the energy ($x$-axis), normalized to the energy gap of the system $\mathcal{E}/\epsilon$.
    The parameter $p$ represents the strength of the dissipation from $p=0$ (no dissipation) to $p=1$ (full relaxation process).
    The blue dots represent the theoretical predictions of the discussed detection scheme.
    The red triangles represent the theoretical predictions of the two-measurement process.
    The quasi-probability distribution functions have been rescaled to plot them with the probability distribution obtained with the TMP \cite{SM}.
    In the 
    experiments we set $\theta=0.7$, $\phi=1.2$ for the initial state and  $\alpha=1$ and $\beta=0.5$ for the system dynamics.}  
    \label{fig:distributions}
\end{figure*} 

We model the dissipation as an amplitude-damping channel \cite{nielsen-chuang_book, Garcia-Perez2020} in which the environment   induces relaxation from the excited to the ground state with probability $p$ \cite{SM}.
The parameter $p$ determines the strength of the dissipation process and it ranges from $p=0$, i.e., no dissipation, to $p=1$, complete relaxation to the ground state.

If the relaxation occurs in the $\{\ket{0}, \ket{1} \}$, i.e., the $\sigma_z$, basis, this corresponds to the transformation
\begin{eqnarray}
 \ket{00} &\rightarrow& \ket{00} \nonumber \\
 \ket{10} &\rightarrow& \sqrt{1-p} \ket{10} + \sqrt{p} \ket{01},
  \label{eq:cold_environment}
\end{eqnarray}
where the first and the second qubits represent the system and the environment, respectively. 
Physically, this describes the emission of an energy quantum from the system, i.e., the relaxation, and the corresponding absorption from the environment.
We assume that the environment is at temperature $T$ and that $k_B T \ll \epsilon$ (where $k_B$ is the Boltzmann constant) so that  the processes in which the system is excited by the interaction with the environment are exponentially suppressed and are neglected in Eq. (\ref{eq:cold_environment}).
The logical operator that mimics the transformation (\ref{eq:cold_environment}) is denoted as $R_z$  (see figure \ref{fig:gate_sequence} and \cite{SM}).
When the relaxation occurs in the $\sigma_x$ basis, i.e., during the first part of the evolution, the transformation is analogous to  (\ref{eq:cold_environment}) and is implemented with a simular operator denoted by the $R_x$ operator \cite{SM}.

To realize the effect of this engineered environment with the $R_x$ and $R_z$ operators (see figure \ref{fig:gate_sequence}), we need two additional two-level systems. On a NISQ (Noisy Intermediate-Scale Quantum) device this translates into two additional qubits \cite{SM}.


In addition, we assume that the unitary evolution occurs on shorter time scales with respect to the relaxation time scale. Hence, we can imagine the system evolution as a sequence of unitary dynamics followed by relaxation.

The building block of the measurement schemes is the operator sequence $U_{\chi, t_i} ~U~ U_{-\chi, t_j}$ where $U$ is a unitary operator acting on the system.
As discussed in Ref. \cite{solinas2015fulldistribution}, this allows us to have information about the {\it system energy changes in the time interval $t_i-t_j$}.
Here, it is of fundamental importance to note that work and heat are associated to different kinds of dynamical evolution \cite{solinas2013work}.
The work is associated to the changes in the Hamiltonian while the heat is associated to the change of the system state when the Hamiltonian is constant.

Since the system Hamiltonian changes only at time $t_1$, to measure the work we couple the system and the detector shortly before and after $t_1$ (Fig. \ref{fig:gate_sequence}).

In the present 
experiment, the (dissipative) dynamics of the system is given by the operator $U_{diss} = R_z~U_z~R_x~U_x$.
The evolution generated by the latter operator can be divided in two parts, $R_x~U_x$ and $R_z~U_z$, where the system Hamiltonian can be considered constant.
Before and after the time $t_1$, when the Hamiltonian is constant, the change in the system energy can be associated to the action of the environment and, thus, interpreted as dissipated heat.
Analogously, we follow the scheme in Fig. \ref{fig:gate_sequence} to store information about the dissipated heat in the $\sigma_x$ and $\sigma_z$ system basis, respectively.
Note that with this coupling scheme, we obtain information about the heat, i.e., the dissipated energy, supplied {\it to the environment} \cite{SM}. If we are interested in the variation of the internal energy of the system, we need to couple the system and the detector at the beginning and at the end of the evolution only (see Fig. \ref{fig:gate_sequence}).

We would like to stress that the separation between the unitary and dissipative dynamics, i.e., between interval in which the work is done and the heat is dissipated, has the only purpose of simplifying the discussion and the implementation  on the NISQ device.
As discussed in Ref. \cite{solinas2015fulldistribution}, the approach works also in more complex situations.
The only constraint is that we must be able to couple the system and the detector on timescale $\Delta t$ such that $\Delta t \ll \Time$.
Under this condition, in a single interval the system Hamiltonian can be considered constant and the the system dynamics (if present) is associated to the effect of the environment and to the dissipated heat.

Coming back to the discussed case, the physical observable measured in the 
experiments is the phase accumulated in the detector using the different schemes.
For a given system-detector coupling strength $\chi$, it reads \cite{solinas2015fulldistribution,solinas2016probing,SolinasPRA2017}
\begin{equation}
  \GFuncF = \frac{_D\matrixel{0 }{\rho_D(\Time) }{1 }_D}{_D\matrixel{0 }{\rho^0_D }{1 }_D } = \Trace_{S,E} \Big[ \mathcal{U}_{\chi,\mathcal{F}} \rho^0_S \mathcal{U}_{-\chi,\mathcal{F}}^\dagger\Big].
  \label{eq:CGF}
\end{equation}
where $\mathcal{F} = \Delta U, W, Q$, $\rho(\Time)$, $\rho_S(\Time) = \Trace_{D,E} [\rho(\Time)] $ and $\rho_D(\Time) = \Trace_{S,E} [\rho(\Time)] $ are the total, the final system and detector density operators, respectively;  $\mathcal{U}_{\pm \chi,\mathcal{F}}$ represents the full (system, detector and environments) evolution \cite{solinas2015fulldistribution, SM}.

By changing the system-detector coupling strength $\chi$, we obtain the quasi-characteristic function (QCGF) that is related to quasi-moments of the observable $\mathcal{F}$ \cite{clerk2011full,solinas2015fulldistribution,solinas2016probing,SolinasPRA2017}.
In particular, the first derivative of $\GFuncF$ with respect to $\chi$ gives the average value of $\mathcal{F}$.
For example, by considering $\mathcal{F} = \Delta U$, i.e., first coupling scheme in Fig. \ref{fig:gate_sequence}, by direct calculation, it can be shown that $d \GFuncU/d\chi \big|_{\chi=0}= \average{H_S(\Time)- H_S(0)} = \Delta U$ \cite{clerk2011full,solinas2015fulldistribution,solinas2016probing,SolinasPRA2017}.

The Fourier Transform of the QCGF allows us to obtain a quasi-probability distribution function:
 $\Prob(\mathcal{F}) = \int d \chi \GFuncF e^{ i \chi \mathcal{F}}$.
By using the prefix "quasi", we stress that the above quantities are not obtained by direct physical measurements but derived from the analysis of the detector phase measurements.
Indeed, $\Prob(\mathcal{F})$ presents negative probability regions \cite{hofer2016, solinas2015fulldistribution}
that are related to the violation of the Leggett-Garg inequalities \cite{Leggett1985, hofer2016, solinas2015fulldistribution, clerk2011full, Lostaglio2018, Levy2020} and can therefore be seen as a signature of the quantum properties of the work.

\section{Results}
\label{sec:results}

The main results of the 
experiments on the IBMQ device are shown in Fig. \ref{fig:distributions} (see \cite{SM} for the details about the implementation) for different the dissipative parameters $p$.
When no relaxation is present, i.e., $p=0$, no heat is dissipated and the heat distributions is peaked around $Q=0$.
The internal energy and work QPDFs are identical and show the classical energy peaks at $\mathcal{E}/\epsilon = \pm 1, 0$. 
However, there are also quantum energy peaks at $\mathcal{E}/\epsilon = \pm 1/2$ corresponding to the exchange of half of an energy quantum, as predicted by the theory \cite{solinas2015fulldistribution,solinas2016probing,SolinasPRA2017}.
More importantly, in this regions the probability density distribution can be negative, thus signaling a pure quantum effect related to the violation of the Leggett-Garg inequalities \cite{Leggett1985, solinas2015fulldistribution}.

The 
experimental data and QPDFs are in excellent agreement with the theoretical predictions, which are presented as blue dots.
The red dots represent the expected values obtained from a numerical simulation of the TMP, where the system energy is initially projectively measured.
Clearly, the QPDFs not only contain all the information and reproduces the TMP distributions but they allow us to determine the presence of additional quantum interference effects highlighted by the half-quantum energy exchanges \cite{solinas2015fulldistribution,solinas2016probing}.

For intermediate values of the system-environment coupling strength ($p=0.5$ in Fig. \ref{fig:distributions}), the QPDFs of $W$ and $\Delta U$ have some notable differences but the features described above persist. However, in this case some heat is being dissipated by the system as pointed out by the presence of a peak at energy $Q=\mathcal{E}/\epsilon = 1$. 
Notice that, while the internal energy is bounded between $\pm 1$, the dissipated heat is not.
Because of the chosen dynamics, the system can dissipate energy at two times (formally when we apply $R_x$ and $R_y$) corresponding to a maximum energy exchange of two energy quanta.

The strong dissipation case $p=1$ has two interesting features.
First, as expected, the peaks of the heat distribution are evidently signaling an increased dissipation.
Second, the $W$ and $\Delta U$ distribution are now both positively defined and no quantum energy exchange at $\mathcal{E}/\epsilon = \pm 1/2$ is present.
The disappearance of the quantum region in the QPDFs signals the emergence of the classical limit due to the interaction of the system with an environment.

As a side remark, we note that the $Q$ distribution is always classical, i.e., positively defined and with no half-quantum energy exchange.
This is a feature we expect from a large Markovian environment that is always in equilibrium and, in this sense, a classical environment.

\begin{figure}
    \begin{center}
    \includegraphics[scale=.55]{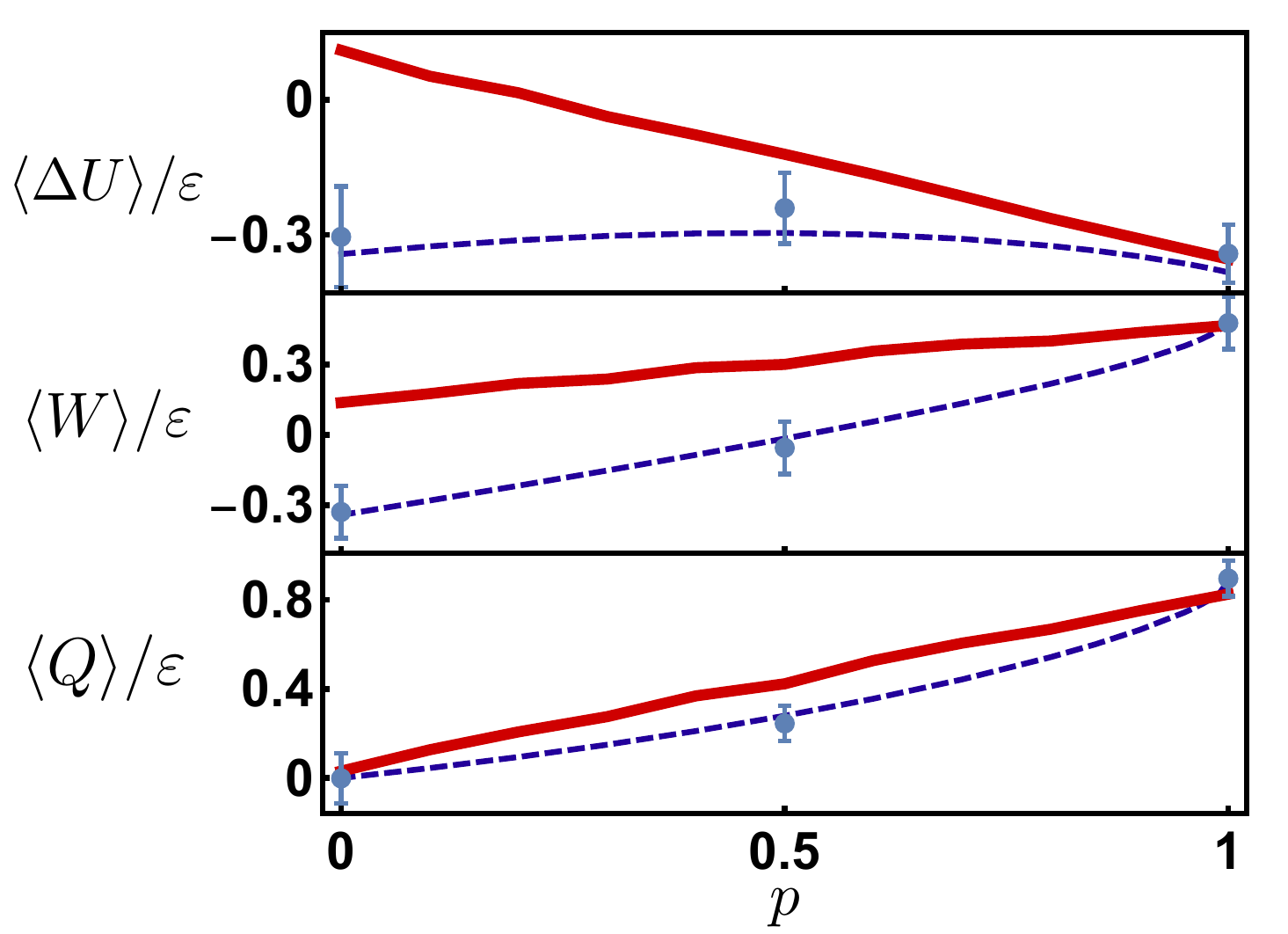}
       \end{center}
    \caption{The average values of the variation of internal energy, work and dissipated heat normalized to $\epsilon$ as a function of the dissipative parameter $p$.
    The blue dots represent the 
    experimental data with the corresponding error bars (see \cite{SM} for details).
    The dashed blue curve is the theoretical prediction of the discussed approach.
    The red solid curve is the prediction of the TMP obtained by simulations with the IBMQ computers. The error in these simulations is within the curve thickness.
    As can be seen, for strong dissipation the three curves converge demonstrating that the quantum features are destroyed and the classical limit is reached.
     }  
    \label{fig:averages}
\end{figure}

These results and trends are confirmed by the behavior of the average values shown in Fig. \ref{fig:averages} as a function of the dissipative parameter $p$.
The 
experimental points (in light blue) are in good agreement with the expected theoretical prediction (blue curve).
For no or weak dissipation ($p=0$ and $p=0.5$), these are different from the one predicted by the TMP (red curve).
The difference lies in the initial coherences and interference effects that are preserved with the present approach.
For strong dissipation ($p=1$), the averages obtained with different approaches coincide.
Since the main effects of a strong dissipation is to destroy the quantum coherences and make the evolution classical, this is another manifestation of the emergence of the classical limit that coincides with the TMP results.

All the averages satisfy the energy conservation law $\average{\Delta U} + \average{Q} - \average{W}=0$. In particular, we find from the 
experimental data that the energy conservation law is satisfied within a the experimental error \cite{SM}.

\section{Conclusions}
\label{sec:conclusions}

The present approach has several advantages with respect one presented in Refs \cite{campisi2011colloquium, gasparinetti2014heat}.
First, it allows us to obtain the dissipated heat by acting on the system degrees of freedom. This is an important difference with respect to the theoretical proposal to measure the variation of the energy of the environment, which is practically unrealizable since it would require an insurmountable number of measurements.
Other viable schemes have been designed in order to measure the dissipated energy quanta \cite{pekola2013calorimetric, Gasparinetti2015, Viisanen_2015, Saira2016, Vesterinen2017}. However, these can be used only in specific physical systems while our scheme is system-independent and, thus, can be used with any quantum system.

More importantly, the information about the classical TMP distributions is included in the more general QPDFs that, at the same time, contains much more valuable information about the quantumness of the process.
In Wigner's spirit \cite{WignerPhysRev1932} these can be see as quantum correction to an underlying classical process.
The Wigner function  has become an invaluable tool to understand quantum phenomena \cite{Bertet2002, Deleglise2008}.
Analogously, the presented approach could shed light on the energy exchanges allowing us to identify their quantum features. This is the first step toward understanding if and when quantum energy exchanges can be more efficient that classical ones and exploit their quantum advantage.


\section*{Acknowledgements}
The authors would like to thank E. De Vito, S. Gasparinetti and A. Toigo for fruitful discussions.
PS and NZ acknowledge financial support from INFN.
The views expressed are those of the author and do not reflect the official policy or position of Q-CTRL.


\newpage

\pagebreak
\widetext
\begin{center}
\textbf{\large Supplemental Information}
\end{center}
\setcounter{equation}{0}
\setcounter{figure}{0}
\setcounter{table}{0}
\setcounter{page}{1}
\makeatletter
\renewcommand{\theequation}{S\arabic{equation}}
\renewcommand{\thefigure}{S\arabic{figure}}
\renewcommand{\bibnumfmt}[1]{[S#1]}
\renewcommand{\citenumfont}[1]{S#1}

\section{Operator decomposition in IBMQ fundamental gates}
\label{app:implementation}

To implement the protocol described in the main text on the IBM quantum computer we need the Hadamard gate, the phase gate $u_1(\theta)$ and the controlled-NOT gate \cite{IBM_docs}.
The first two read in the  $\{ \ket{0}, \ket{1} \} $ basis
\begin{equation}
H = \frac{1}{\sqrt{2}} \left(
\begin{array}{cc}
 1 & 1 \\
 1 & -1 \\
\end{array}
\right)
\end{equation}
and 
\begin{equation}
u_1(\theta) = \left(
\begin{array}{cc}
 1 & 0 \\
 0 & e^{i \theta } \\
\end{array}
\right).
\end{equation}

The initial state of the system $\ket{0}_S$ can be transformed in $\ket{\psi_0} = \cos \frac{\theta}{2} \ket{0}_S+ \sin \frac{\theta}{2} e^{i \phi}\ket{1}_S$ by the application of the operators 
\begin{equation}
 U_{in} =  u_1\left(\varphi + \frac{\pi}{2}\right) ~H ~u_1(\theta)~H.
 \label{app_eq:initialization}
\end{equation}
The dynamical operator acting on the system is $U_S (\alpha , \beta) = \exp\{- i \beta \sigma_z\} \exp\{- i \alpha \sigma_x \}$ which can be obtained (a part from a irrelevant $\exp\{ i (\beta+ \alpha)\}$ factor) from the gate sequence $U_S (\alpha , \beta) = u_1(\beta) ~H~u_1(\alpha)~H$.
In the quantum simulated experiments and theoretical simulations, we have set $\theta=0.7$, $\phi=1.2$ for the initial state and  $\alpha=1$ and $\beta=0.5$ for the system dynamics.

The system-detector coupling prototype operator reads 
\begin{equation}
 U_\chi = \left(
\begin{array}{cccc}
 e^{i \chi } & 0 & 0 & 0 \\
 0 & e^{-i \chi } & 0 & 0 \\
 0 & 0 & e^{-i \chi } & 0 \\
 0 & 0 & 0 & e^{i \chi } \\
\end{array}
\right)
\end{equation}
in the $\{ \ket{00}, \ket{01}, \ket{10}, \ket{11}\}$ basis.

We denote with $\CNOT{i}{j}$ the controlled-NOT gates that has the $i$-th qubit as the control and $j$ has the target qubit  \cite{IBM_docs}.
If the coupled qubits are the system ($i=1$) and the detector one ($j=2$), the corresponding $\CNOT{1}{2}$ gate in the same basis reads
\begin{equation}
 \CNOT{1}{2}= \left(
\begin{array}{cccc}
 1 & 0 & 0 & 0 \\
 0 & 1 & 0 & 0 \\
 0 & 0 & 0 & 1 \\
 0 & 0 & 1 & 0 \\
\end{array}
\right).
\end{equation}
When $H_S = \epsilon \sigma_z$,  $U_\chi = \exp\{ i \chi \sigma_z \otimes \sigma_z\}$ and it can be implemented as 
\begin{equation}
 U_{\chi,z} = \CNOT{1}{2}~\left [  \Idoperator \otimes u_1(-2 \chi) \right]~\CNOT{1}{2}
 \label{app_eq:U_chi_z}
\end{equation}
(a part a $\exp\{i \chi \}$ factor).

We also need another system-detector coupling operator when $H_S = \epsilon \sigma_x$; that is $U_\chi = \exp\{ i \chi \sigma_x \otimes \sigma_z\}$.
The logical gate sequence reads
\begin{equation}
 U_{\chi,x} = H~\CNOT{1}{2}~\left [  \Idoperator \otimes u_1(-2 \chi) \right]~\CNOT{1}{2}~ H.
  \label{app_eq:U_chi_x}
\end{equation}

Finally, we need to measure the phase accumulated by the detector due to the interaction with the system.
The real and imaginary part of the off-diagonal matrix elements are measured in separate quantum simulated experiments. 
For the real part, this is done by applying the Hadamard operator followed by a measurement.
For the imaginary part, applying the operator $\exp\{- i \pi \sigma_x/4\}$ followed by a measurement.
The second operator can be implemented by the use of the IBMQ $u_2 (\pi/2,-\pi/2)$ gate that, in the $\{ \ket{0}, \ket{1} \} $ basis, is \cite{IBM_docs}
\begin{equation}
u_2 (a,-a) = \frac{1}{\sqrt{2}} \left(
\begin{array}{cc}
 1 & -e^{-i a} \\
 e^{i a} & 1 \\
\end{array}
\right).
\end{equation}


\section{Engineered environment}
\label{app:engineered_environment}

We make the working hypothesis that the main source of decoherence is the relaxation of the system qubit from the excited to the ground state.
To describe this type of relaxation we use an engineered amplitude-damping channel \cite{nielsen-chuang_book, Garcia-Perez2020}.
This allows us to study the work and heat distribution at different level of environmental coupling.

The system-environment qubit register is composed by a system qubit $q_s$ and an environment qubit $q_{env}$ so that the full quantum state describing the register is $\ket{q_s, q_{env}}$.
The transformation we implement is
\begin{eqnarray}
 \ket{00} &\rightarrow& \ket{00} \nonumber \\
 \ket{01} &\rightarrow& \sqrt{1-p} \ket{01} - \sqrt{p} \ket{10}  \nonumber \\
 \ket{10} &\rightarrow& \sqrt{1-p} \ket{10} + \sqrt{p} \ket{01} \nonumber \\
 \ket{11} &\rightarrow& \ket{11}.
 \label{app_eq:amplitude_damping_channel}
\end{eqnarray}
It describes the the exchange of an energy quantum between the system and the environment and the parameter $p$ is to the probability for the transition $\ket{01} \leftrightarrow \ket{10}$.

In the cold environment limit, i.e., $k_B T \ll \epsilon$ where $T$ is the bath temperature, the system cannot be excited by the environment but can only relax to the ground state.
In this case, assuming that the environment is in the ground state, the above transformation is 
\begin{eqnarray}
 \ket{00} &\rightarrow& \ket{00} \nonumber \\
 \ket{10} &\rightarrow& \sqrt{1-p} \ket{10} + \sqrt{p} \ket{01}.
  \label{app_eq:cold_environment}
\end{eqnarray}

If the system is initially in the state $\ket{\psi_s}  = \alpha \ket{0} + \beta e^{i \varphi}\ket{1}$, the total initial state evolves into
\begin{equation}
 \ket{\psi_{in}} = \alpha \ket{00} + \beta e^{i \varphi}\ket{10}  \rightarrow \ket{\psi_{fin}} = \alpha \ket{00} + \beta e^{i \varphi} ( \sqrt{1-p} \ket{10} + \sqrt{p} \ket{01})
\end{equation}
which corresponds in terms of the system density matrix reduced dynamics (in the $\{ \ket{0}, \ket{1} \} $ basis) to
\begin{equation}
\rho_{S,in} = 
\left(
\begin{array}{cc}
 \alpha^2 & \alpha \beta e^{-i \varphi} \\
 \alpha \beta e^{i \varphi}  & \beta^2 \\
\end{array}
\right) 
\rightarrow
\rho_{S,fin} =
\left(
\begin{array}{cc}
 \alpha^2 + p \beta^2  & \alpha \beta e^{-i \varphi} \sqrt{1-p}\\
 \alpha \beta e^{i \varphi}  \sqrt{1-p} & (1-p) \beta^2 \\\end{array}
\right).
\label{eq_app:amplitude-damping}
\end{equation}
In this case, $p$ represent the probability for the excited state to relax to the ground state and, therefore, it characterises the strength of the environmental effects.
As we can see, for $p=0$ there is not relaxation and for $p=1$ the relaxation is complete, i.e., the system ends up into the $\ket{0}$ state.

The same results can be obtained with the the operator sum-representation \cite{nielsen-chuang_book, Garcia-Perez2020}.
The operator elements in the system $\{\ket{0}, \ket{1} \}$ basis are 
\begin{eqnarray}
M^0 &=& 
\left(
\begin{array}{cc}
 1 & 0 \\
 0 &  \sqrt{1-p} \\
\end{array}
\right) \nonumber \\
M^1 &=&
\left(
\begin{array}{cc}
 0  & \sqrt{p} \\
 0& 0 \\
 \end{array}
\right).
\label{eq_app:operator_elements}
\end{eqnarray}
The final system density operator is $\rho_{S,fin} = M^0 \rho_{S,in} M^0 + M^1 \rho_{S,in} M^1$. 
By direct calculation we obtain the result in Eq. (\ref{eq_app:amplitude-damping}).

\subsection{Implementation in Qiskit}

For the implementation in Qiskit, we use the ${\rm C}u_{3,i,j}(\theta)$ gate that induces a controlled rotation of an angle $\theta$ on the $j$-th qubit if the $i$-th qubit is active \cite{IBM_docs}.
The rotation on the target qubit generates $\ket{0} \rightarrow \cos \frac{\theta}{2} \ket{0} + \sin \frac{\theta}{2} \ket{1}$ and $\ket{1} \rightarrow \cos \frac{\theta}{2} \ket{1} - \sin \frac{\theta}{2} \ket{0}$.
The ${\rm C}u_{3,i,j}(\theta)$ acting on a generic $\ket{i,j}$ state produces
\begin{eqnarray}
 \ket{00} &\rightarrow& \ket{00} \nonumber \\
 \ket{01} &\rightarrow&  \ket{01}  \nonumber \\
 \ket{10} &\rightarrow& \cos \frac{\theta}{2} \ket{10} + \sin \frac{\theta}{2} \ket{11} \nonumber \\
 \ket{11} &\rightarrow& - \sin \frac{\theta}{2}  \ket{10} + \cos \frac{\theta}{2} \ket{11}.
\end{eqnarray} 

In our case, the working space is composed by the system qubit $q_s$ and by one environmental qubit $q_{env}$ so that the full state reads $\ket{q_s, q_{env}}$.
The logical gate sequence needed is
\begin{equation}
  {\rm Relaxation}_{q_s,q_{env}} = \CNOT{q_{env}}{q_s}~{\rm C}u_{3,q_s,q_{env}}(\theta) ~\CNOT{q_{env}}{q_s}
  \label{app_eq:relaxation_z}
\end{equation}
(Notice the inversion in the controlled operations, i.e., for the $\CNOT{q_{env}}{q_s}$ gates the control is on the environment qubit and the target is the system qubit while for the ${\rm C}u$ operation the control is on the system qubit while the rotation is applied to the environment qubit).
The total transformation on a $\ket{q_s, q_{env}}$ state is
\begin{eqnarray}
 \ket{00} &\rightarrow& \ket{00} \nonumber \\
 \ket{01} &\rightarrow& - \sin \frac{\theta}{2}  \ket{10} + \cos \frac{\theta}{2} \ket{11}  \nonumber \\
 \ket{10} &\rightarrow& \cos \frac{\theta}{2} \ket{10} + \sin \frac{\theta}{2} \ket{11} \nonumber \\
 \ket{11} &\rightarrow&  \ket{11} .
\end{eqnarray} 
This is reduced to the amplitude-damping channel [Eq. (\ref{app_eq:amplitude_damping_channel})] when $\theta = 2 \arctan[ \sqrt{1-p}]$ and to Eq. (\ref{app_eq:cold_environment}) for a cold environment.
Therefore, the gate sequence in Eq. (\ref{app_eq:relaxation_z}) describes the relaxation of the system qubit in the  {\it in the $\{\ket{0}, \ket{1}\}$ basis}, i.e., when $H_S = \epsilon \sigma_z$.

The relaxation in the $\{\ket{+}, \ket{-}\}$ basis has the same structure with a Hadamard rotation on the system qubit, i.e., 
\begin{equation}
  {\rm RelaxationX}_{q_s,q_{env}} = H_{q_s}~\CNOT{q_{env}}{q_s}~{\rm C}u_{3,q_s,q_{env}}(\theta)~\CNOT{q_{env}}{q_s}~H_{q_s}.
  \label{app_eq:relaxation_x}
\end{equation}


\begin{figure}
    \begin{center}
    \includegraphics[scale=.5]{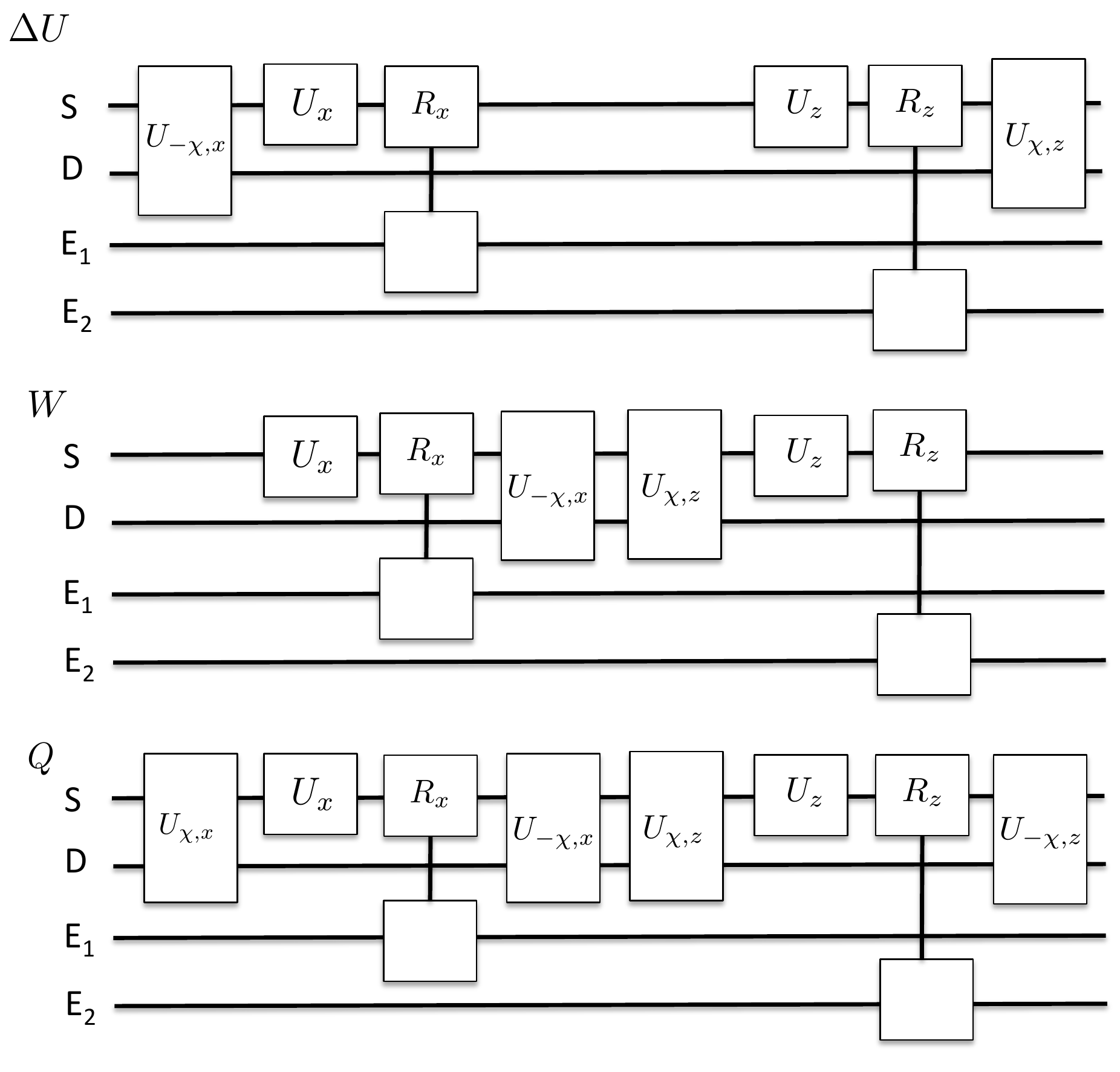}
       \end{center}
    \caption{ Logical gate sequence for the measurement of the average values of work $W$, heat $Q$ and internal energy variation $\Delta U$.
    The gate sequenced needed to initialize the system and to measure the detector phase are not shown.
     }  
    \label{fig:full_implementation}
\end{figure} 

\section{Full dynamics implementation}
\label{app:full_implementation}

The operation sequences needed to build the quasi-probability distribution functions of the work, heat and internal energy variation are discussed in details in Refs. \cite{solinas2015fulldistribution,solinas2016probing} and depicted in Figure $1$ of the main text.

However, for the sake of clarity, in Fig. \ref{fig:full_implementation} we explicitly show the logical gate sequence used to simulate the driven and dissipative dynamics on the IBMQ device.
Here, the logical gate sequence in Eq. (\ref{app_eq:initialization}) used to initialize the system and the ones used to measure the real and imaginary part of the detector phase are not shown.

The $U_x$ and $U_z$ (shown as red and light-blue filled squares in the main text) are the external drive dynamics discussed above. The rectangular boxes  represent the system-detector coupling operation $U_{\pm \chi, x}$ and $U_{\pm \chi, z}$.
In the main text, they are associated to boxes with $i)$ dashed red lines from up-left to down-right ($U_{- \chi, x}$), $ii)$ dashed red lines from up-right to down-left ($U_{ \chi, x}$), $iii)$  dashed blue lines from up-left to down-right ($U_{- \chi, z}$) and $iv)$  dashed blue up-right to down-left $U_{ \chi, z}$.
Their implementations in terms of logical gate sequenced are given in Eqs. (\ref{app_eq:U_chi_x}) and (\ref{app_eq:U_chi_z}), respectively.

The dissipative evolution is denoted by $R_x$ and $R_z$ which are obtained with the logical gates sequence (\ref{app_eq:relaxation_x}) and (\ref{app_eq:relaxation_z}).
In the main text they were shown as a dark red ($R_x$) and a dark blue ($R_z$) circle.

Depending on the physical observable, we  fix the value of the system-detector coupling parameter $\chi$, implement the corresponding logical gate sequence and perform a measure of the detector phase (not shown in the scheme).
By repeating this scheme changing $\chi$, we are able to construct the quasi-characteristic generating function and through a Fourier transform the quasi-probability density function.


\section{Data analysis and presentation}
\label{app:data_analysis}

The quantum simulated experimental data are obtained from the IBM quantum computer ibmq$\_$vigo device \cite{IBM_docs}.
For the physical observable $\Delta U$, $W$ and $Q$, we have implemented the corresponding gate sequence presented in Fig. $1$ of the main text.

We have taken the system-detector coupling parameter $0\leq \chi \leq \chi_{max} = 100$ with a discretization of $\Delta \chi = 0.1$.
For each value of $\chi_k = k \Delta \chi$ (with $0 \leq k \leq 1000$), we run $8000$ quantum simulated experiments, determine the phase accumulated in the detector qubit and calculate $\GFuncF$ as in Eq. $1$ of the main text and the implementation discussed above.
The noise level in the quantum simulated experiments on the IBM quantum computers can change depending on different protocol implementations or can increase in time (the quantum computers need periodic recalibrations).
For these reasons, whenever high noise level was observed, we have rerun the experiments (up to $24000$ experiments for every $\chi_k$) and taken the average $\GFuncF$.
We have used the function symmetries, $\re[\mathcal{G}_{\chi}] = \re[\mathcal{G}_{-\chi}]$ and $ \im[\mathcal{G}_{-\chi}] = -\im[\mathcal{G}_{\chi}] $, to extend the function in the interval $- \chi_{max} \leq \chi \leq \chi_{max}$.
From $\GFuncF$ we have calculated the corresponding quasi-probability distribution functions as $\Prob(\mathcal{F}) = \int_{-\chi_{max}}^{\chi_{max}} d \chi \GFuncF e^{ i \chi \mathcal{F}}$. 

The quantum simulated experiments on IBM quantum computer are flanked by numerical simulations.
These are obtained simulating the system dynamics with the same parameter and precision, i.e., the same $ \chi_{max}$ and $\Delta \chi$.

Finally, the TMP distribution results are obtained by running a numerical simulation of the TMP protocol on the ibmq$\_$vigo device \cite{IBM_docs}. 
Since in this case we perform a direct measurement of the probability distribution, we do not need a detector qubit and we used a three-qubit register: one qubit for the system and two for the environment.

To simulate the initial measurement, we prepare the system in one of the two two eigenstates of $H_S(0)$, i.e., $(\ket{0} \pm \ket{1})/\sqrt{2}$, and then apply the same evolution operators $U_{x,z}$ and relaxation operators $R_{x,z}$ as in the QPDF scheme.
At the end of the simulation, all three qubits are measured. The knowledge of the final measured state 
$\ket{q_s, q_{env_1}, q_{env_2}}$ (in the $\{ \ket{0}, \ket{1} \}$ basis) allows us to determine both the variation of internal energy and the dissipated heat.
In this way, we can calculate the variation of internal energy $\Delta U$ as the difference between the energy of the initial state and the final measured state.
We can also calculate the dissipated heat. Since at the beginning we have $q_{env_1} = q_{env_2}=0$, the normalized dissipated heat is $Q/\epsilon= q_{env_1} + q_{env_2}$.
The work can now be calculated as a {\it derived quantity}, equal to $W = \Delta U + Q$.
The results obtained are weighted with the probability to start from the corresponding eigenstate when the measurement is performed on  $\ket{\psi_0} = \cos \frac{\theta}{2} \ket{0}_S+ \sin \frac{\theta}{2} e^{i \phi}\ket{1}_S$.

To compare the quantum simulated experimental results for the QPDFs with the ones obtained from the simulation of the TMP, we focus only on the classical energy peaks (integer quanta exchange).
These correspond to the case in which the system has no initial coherences and the energy processes becomes classical and, thus, reproduced by the TMP \cite{solinas2015fulldistribution,solinas2016probing}.

The relevant quantities to be compared are the probability to obtain a process with energy $\mathcal{E}$.
For the present protocol, it corresponds to the integral $\int_{\mathcal{E}-\delta}^{\mathcal{E}+\delta} d \mathcal{F} \Prob(\mathcal{F})$, that is the probability to exchange an energy between $\mathcal{E}-\delta$ and $\mathcal{E}+\delta$.
On the contrary, for the TMP we obtain a probability mass functions so the probability to obtain $\mathcal{E}$ is obtained evaluating the distribution at the point.
These statistical distributions are intrinsically different, i.e., the quantum simulated experimental data are continuous (probability distribution) functions while the TMP are discrete functions, so to present them together, we need to renormalize the experimental distributions.

The oscillations in the quantum simulated experimental functions are due to the limitation of the quantum simulated experimental data. 
As $\chi_{max}$ increases, the distributions are more and more peaked around the discrete energy exchange ($\mathcal{E}=\{ 0, \pm 1, \pm 2\}$).
Accordingly, we assume that all the information about the physical processes is in these discrete points.
Then, for every probability distribution, we adopted the following procedure:
\begin{enumerate}
\item We calculate the values of the distribution at the peaks and sum them to obtain a norm:
$\mathcal{N} = \sum_{\mathcal{E}} \Prob(\mathcal{\mathcal{E}})$.
\item We then rescale the original (quasi-)probability distribution function with respect to this norm:
$\Prob'(\mathcal{F})= \Prob(\mathcal{F})/\mathcal{N}$. 
\end{enumerate}

The final effect of this procedure is to preserve the relative strength of the peaks  give a direct visual comparison between the TMP distribution peaks and the classical peaks of the quasi-probability distributions.


\section{Error calculation }
\label{app:error}

\begin{figure}
    \begin{center}
    \includegraphics[scale=.6]{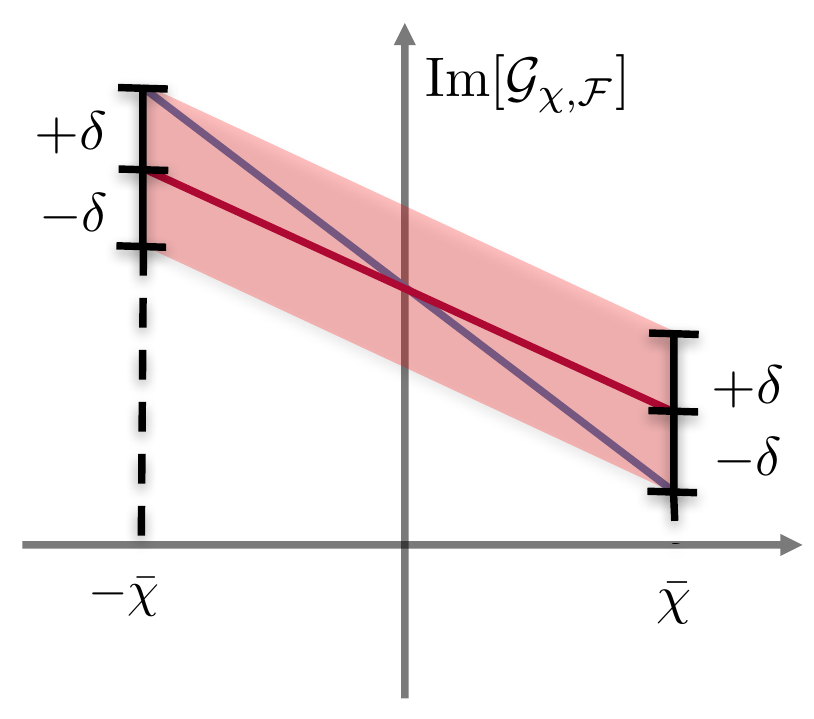}
       \end{center}
    \caption{Geometric calculation of the error. The imaginary part of $\GFuncF$ is linearly approximated around $\chi=0$ and the slope is associated to the average values of the physical observable.    
    The red line represents the behaviour obtained by the quantum simulated experiments.
    At $\bar{\chi} = 0.1$ we have an error $\delta$ determined by the statistical uncertainty in the determination of the ground and excited state populations.
    The error determines a change in the line slope and, consequently, a change in the average values of the physical observables.
} 
    \label{fig:error_figure}
\end{figure}

As discussed above, in the quantum simulated experiments we obtain the real and imaginary part of the phase accumulated by the detector during the evolution by measuring the population of the ground $p_0$ and the excited state $p_1$ of the detector qubit.
These are related to $\GFuncF$  in Eq. $(1)$ of the main text by the expressions
$\re[\GFuncF] = p_0-p_1=2 p_0-1$ for the first measurement scheme, i.e., the one with a $H$ gate, and $\im[\GFuncF] = p_0-p_1=2 p_0-1$ for the second measurement scheme, i.e., the one with a $\exp\{- i \pi \sigma_x/4\}$ gate (we recall that the initial state of the detector has $_D\matrixel{0 }{\rho^0_D }{1 }_D = 1/2$).
From the raw quantum simulated experimental data, we calculate the QCGF and then the QPDF.

The quantum simulated experimental populations $p_0$ and $p_1$ are affected by errors.
Since it is not possible to characterize the error induced by the dynamics, i.e., the sequence of logical gates, we consider only the statistical error due to the final measurement.

The interesting physical quantities are the average values of $\Delta U$, $W$ or $Q$ therefore we only discuss the experimental error on these.
The propagation of the experimental error to the QPDF and then the average values is not straightforward because it requires the definition of a "distance" between the two QPDFs (the measured one and the one with the included errors).
Here, we use a more direct and simple way to calculate the error on the average quantities of interest.

The  $\GFuncF$ is built so that the first derivative (calculated in $\chi=0$) reproduces the average values of the physical observables $\mathcal{F}$ \cite{solinas2015fulldistribution,solinas2016probing,SolinasPRA2017}, i.e.,
\begin{equation}
\average{\mathcal{F}} = -i \frac{d \GFuncF}{ d\chi} \Big | _{\chi=0}.
\label{eq_app:first_moment}
\end{equation}
Note that, given the symmetries of $\GFuncF$, only the imaginary part contributes to the first moment.

As shown in \cite{solinas2015fulldistribution,solinas2016probing,SolinasPRA2017}, the accumulated phase on the detector is $\exp{ \{ i \chi (\epsilon_i - \epsilon_j) \} }$ where $\epsilon_i - \epsilon_j$ is the normalized energy variation during the process.
In our case, $\epsilon_i - \epsilon_j = 0, \pm 1$ for $\Delta U$ and $W$ and $\epsilon_i - \epsilon_j = 0, \pm 1, \pm 2$ for $Q$. Therefore, the imaginary part of $\GFuncF$ behaves as $\sin [\chi (\epsilon_i - \epsilon_j)]$ and can be linearly approximated for $\chi \ll 1/(\epsilon_i - \epsilon_j) \leq 1/2$.
In the linear approximation (see in Figure \ref{fig:error_figure}), $\average{\mathcal{F}}$ is the slope of the line $\mathcal{G}_0 + \average{\mathcal{F}} \chi$.
In this geometric interpretation, the error on the populations $p_0$ and $p_1$ determine an error $\delta$  in $\im[\GFuncF]$ resulting, eventually, in a change of slope and the average value of the physical observable (see Figure \ref{fig:error_figure}).

More quantitatively, the change in the slope due to $\delta$ at point $\chi= \bar{\chi}$ can be determined by geometric means and reads
\begin{equation}
 \Delta \average{\mathcal{F}} = \Big | \frac{\delta}{\bar{\chi}} \Big |.
 \label{eq_app:first_moment_error}
\end{equation}

For every value $\chi$, we performed $N=8000$ quantum simulated experiments to determine the imaginary part of $\GFuncF$.
Since the experiments are independent, the error on the determination of the populations $p_i$ can be estimated to be $ 1/\sqrt{N}$ (a more precise description of the error does not change significantly the results).

The smallest positive value of $\chi$ at which the $\GFuncF$ was estimated is $\bar{\chi} = 0.1$.
We have repeated the quantum simulated experiments $3 N$ times for $\mathcal{G}_{\chi,\Delta U}$ and $p=1$, $2 N$ times for $\mathcal{G}_{\chi,\Delta U}$ and $p=0.5$ and  $\mathcal{G}_{\chi,Q}$ and $p=0.5, 1$.
For all the other quantum simulated experiments we have performed $N$ runs.
These data with $\delta = 1/\sqrt{N}$ allow us to calculate the errors for the physical averages presented in the main text.

The theoretical prediction of the averages have been calculated with numerical simulations.
We have plotted the averages calculated directly as the derivative of the QCGFs.
The theoretical calculation through the QPDFs give the same results.

Finally, the averages of $\Delta U$, $W$ and $Q$ for the TMP, are obtained by numerical simulation with the IBMQ computers.
The statistical error in this case is $1/\sqrt{N_{TMP}}$ where $N_{TMP}$ are the number of simulated experiments performed.
In this case, the error bar are so small that are not visible in the plot.

\section{Notation of the heat sign}
\label{app:heat_sign}

The coupling sequence $U_{\chi, t_i} ~U~ U_{-\chi, t_j}$ where $U_{\pm \chi,t} = \exp\{ i \chi H_S(t) \otimes H_D \}$, allows us to determine the variation of the internal energy of the system, i.e., $\Delta U = \epsilon_f - \epsilon_i$ \cite{solinas2015fulldistribution,solinas2016probing}.
This is the energy {\it supplied to the system}; if the system absorbs energy $\Delta U>0$ and if emits energy, $\Delta U<0$.
This can be seen directly by calculating the CGF and the average value as discussed in Ref. \cite{solinas2015fulldistribution}.

On the contrary the transformation  $U_{-\chi,t_i} ~U_{diss}~ U_{\chi, t_j}$ 
measures the energy  {\it supplied by the system} to the environment.
In other words we are "measuring"  $q = \epsilon_i - \epsilon_f$.
Therefore, if $q=\epsilon_i - \epsilon_f>0$ the system emits an energy quantum decreasing its energy while the environment increases its energy absorbing an energy quantum.
If $q=\epsilon_i - \epsilon_f<0$ the system absorbs an energy quantum increasing its energy while the environment decreases its energy emitting an energy quantum.

Notice that this is the opposite notation with respect to the usual one where the $q$ is the energy {\it supplied to the system} as heat.
As a consequence, the energy conservation law takes the form $ \average{\Delta U} + \average{Q}-\average{W} =0$.


\end{document}